\documentclass[hypertex,letterpaper,twocolumn,showpacs,preprintnumbers,amsmath,amssymb,floatfix]{revtex4}

\usepackage{graphicx}
\usepackage{dcolumn}
\usepackage{bm}
\usepackage[colorlinks=true]{hyperref}

\newcommand{\fr}[2]{\textstyle{\frac{#1}{#2}}}
\newcommand{\tin}[1]{{\mbox{\tiny $ #1 $}}}
\newcommand{\mui}{\mu_{\mbox{\tiny I}}}
\begin{document}

\preprint{}

\title{Thermal pion masses in the second phase: $|\mu_{I}| >m_{\pi}$ }

\author{M. Loewe}
\author{C. Villavicencio}%
\affiliation{%
 Facultad de F\'{\i}sica,
 Pontificia Universidad Cat\'olica de Chile,
 Casilla 306, Santiago 22, Chile}%

\begin{abstract}
Density and thermal corrections to the mass of the pions are
studied in the framework of the $SU(2)$ low energy effective
chiral lagrangian, in terms of the isospin chemical potential
$\mu_\tin{I}$. We concentrate the discussion in the region where
the isospin chemical potential (absolute value) becomes bigger
than the pion mass at zero temperature and density, i.e. in the
phase where the condensed $\pi^-$ phase appears (for negative
chemical potential). We are able to calculate the thermal and
density evolution of masses in the limits where $|\mu_{I}|\gg m$
and where $|\mu_{I}|\ \gtrsim m$. We also identified the phase
transition curve.
\end{abstract}


\maketitle

In this paper we continue with the discussion of the thermal and
density pion properties that we started in 
\cite{Loewe:2002tw,Loewe:2003dq} now in the
so called second phase where the absolute value of the isospin
chemical potential becomes bigger than the tree-level pion mass at
zero temperature and density. In the first phase, unless we go to
the chiral limit, the Isospin symmetry $SU(2)$ is broken due to
mass effects. This effect, as soon as we take $|\mu _{I}|>m_{\pi}$
becomes extremely important since one of the charged pions
condense. Here we will use the following definition of the isospin
chemical potential $\mu_\tin{I}\equiv\mu_u -\mu_d$, being
$\mu_{u,d}$ the chemical potential corresponding to the baryon
number of the up and down quarks. In the case of pions, the
isospin chemical potential will correspond essentially to the
difference between the number of positive and negative charged
pions.

We recall that pion masses  at finite temperature $m_{\pi}(T)$
have been studied in a variety of frameworks, such as thermal
QCD-Sum Rules 
\cite{Dominguez:1996kf}, Chiral Perturbation Theory ($\chi$PT)
(low temperature expansion) 
\cite{Gasser:1986vb}, the Linear Sigma Model
\cite{Larsen:1985ei,Contreras:1989gi}, the Mean Field Approximation 
\cite{Barducci:1991rh}, the Virial Expansion 
\cite{Schenk:1991xe}, etc. In fact, the pion propagation at
finite temperature has been calculated at two loops in the frame
of $\chi$PT 
\cite{Schenk:1993ru,Toublan:1997rr}. In our article, we follow the way of
introducing chemical potentials in $\chi$PT that was presented for
the first time in 
\cite{Kogut:1999iv,Kogut:2000ek}. Even though, these articles
deal with QCD with two colors rather than QCD with three colors,
it is clear that both problems are intimately related.

The introduction of in-medium processes via isospin chemical
potential has been studied at zero temperature 
\cite{Son:2000xc,Son:2000by,Kogut:2001id}
in both phases ($|\mu_\tin{I}|\lessgtr m_\pi$) at tree level. In
medium properties at finite density have been discussed for a
variety of phenomena, as for example, the chiral condensates
\cite{Peng:2003jh,Peng:2003nq}, the anomalous decays of pions and etas 
As the density increases both, the quark condensates and the decay
rates diminish.

Interesting results concerning the structure of the QCD phase
diagram, including temperature effects  have been achieved for the
case when we have simultaneously various baryon chemical potential
and isospin chemical potential simultaneously. Two completely
different approaches have confirmed a qualitative change in the
phase diagram as soon as the isospin chemical potential starts to
grow 
\cite{Toublan:2003tt,Barducci:2003un}.
 These  problems concerning the structure of the phase transition diagram,
 have been also handle in the frame of two color QCD in four dimensions
 \cite{Splittorff:2002xn} as well as in three dimensions 
 \cite{Dunne:2003ji}.
 The  problem with
  baryonic chemical potential has been considered in the frame of $\chi$PT
  \cite{Alvarez-Estrada:1995mh}, and also using the finite pion number chemical potential 
  \cite{Ayala:2002qy}.

Different properties of QCD (or QCD-inspired models) under these
kind of circumstances have also been analyzed in the lattice
approach. For QCD with three and two colors, an extensive work has
been carried out 
\cite{Kogut:2001if,Kogut:2002tm,Kogut:2002zg,Kogut:2002cm,Kogut:2003ju}, in connection with the behavior of
different order parameters in several phase transitions in the
$\mu /T$ plane, as for example, the transition to a diquark phase,
the chiral condensate, etc.

A common feature of all approaches in the second phase is the fact
that the physical pion states do not correspond anymore to the
usual  pion  charged states. In fact as we will see, the
$\pi_\tin{\pm}$ states will mix in a non trivial way. The result
of this mixture produces an extremely cumbersome propagator, a
matrix $2 \times 2$, so we need to define some criteria how to
handle this propagator, according to the value of the chemical
potential, for computing radiative corrections.

This paper is organized as follows: First we briefly present the
chiral lagrangian we will use, specifying the lagrangians of
second and fourth order, in the momentum expansion. Then we
present our criteria to expand the mass corrections in a Taylor
series, in two different limits: a) when the chemical potential is
a little bit bigger than the the pion mass and b) in the region
where the chemical potential is much bigger than the pion mass. In
both limits we will work at the lowest order in our expansion.
This will allow us to make a selection of the vertices that will
appear in the different diagrams. We have to note, however, that
our procedure is only valid for values of the chemical potential
up to the mass of the $\eta $ meson. Beyond this point an $SU(3)$
chiral lagrangian is needed. After the selection of relevant
vertices, we compute the thermal mass corrections from the
propagators at the one loop level. We employ the usual momentum or
energy expansion of the propagators around the tree level mass.
This will allow us to comment on the phase transition due to the
pion condensation and compare with similar results from other
authors, 
\cite{Splittorff:2002xn}. Finally we present our conclusions.

\section{ The chiral lagrangian}

The procedure we will follow is basically the same of 
\cite{Loewe:2002tw,Loewe:2003dq}.
The only modification will be the inclusion of non trivial vacuum
expectation values for the pion fields. Nevertheless, this is not
just a little detail since the discussion becomes much more
technically involved and subtle.

In the low-energy description where only pion degrees of freedom
are relevant, the most general chiral invariant lagrangian at the
second order, ${\cal O}(p^2)$, according to an expansion in powers
of the external momentum is given by
\begin{equation}
    {\cal L}_2  = \frac{f^2}{4}Tr\left[(D_\mu U)^\dag D^\mu U
      +2BM(U^\dag + U)\right]
\end{equation}
with
\begin{eqnarray}
 D_\mu U &=& \partial_\mu U-i\mu_\tin{I}u_\mu[\textstyle{\frac{1}{2}}\tau_3, U]\nonumber\\
 U &=& \bar U^\frac{1}{2}(e^{i\pi^a\tau^a/f})\bar U^\frac{1}{2},
\end{eqnarray}
where $\bar U$ is the vacuum expectation value of the field $U$.
$M=diag(m_u,m_d)$ is  the quark mass matrix and $B$ in the
previous equation is an arbitrary constant which will be fixed
when the mass is identified setting $(m_{u} + m_{d})B =m^{2}$,
where $m$ denotes the bare (tree-level) pion mass. We will use
$m_\pi$ to denote the pion masses after renormalization.

 The most general ${\cal O}(p^4)$ chiral lagrangian has the form
\begin{eqnarray}
 {\cal L}_4 &=&
  \textstyle{\frac{1}{4}}l_1
   \left( Tr\left[(D_\mu U)^\dag D^\mu U\right]\right)^2\nonumber\\
 &&
  +\textstyle{\frac{1}{4}}l_2
   Tr\left[(D_\mu U)^\dag D_\nu U\right]Tr\left[(D^\mu U)^\dag D^\nu U\right]\nonumber\\
 &&
  +\textstyle{\frac{1}{4}}(l_3+l_4)
   B^2\left(Tr\left[M( U^\dag +U)\right]\right)^2\nonumber\\
 &&
  +\textstyle{\frac{1}{4}}l_4
   B Tr\left[(D_\mu U)^\dag D^\mu U\right]Tr\left[M( U^\dag + U) \right]\nonumber\\
 &&
  -\textstyle{\frac{1}{4}}l_7
   B^2\left(Tr\left[M( U^\dag-U)\right]\right)^2\nonumber\\
 && + \mbox{constants}.
\end{eqnarray}

We have used the  constants $l_i$ introduced by Gasser and
Leutwyler, 
\cite{Gasser:1983yg} for an $SU(2)$ Lagrangian. They have to be
determined experimentally and are also tabulated in several
articles and books. This Lagrangian includes the chemical
potentials in the covariant derivatives and in the expectation
value of the $\bar U$ matrix. The constants $l_i$ include
divergent corrections which allow to cancel the divergences from
loops corrections
\begin{equation}
l_i(\Lambda)=\frac{\gamma_i}{32\pi^2}\left[\bar l_i-\lambda
  +\ln\left(\frac{m^2}{\Lambda^2}\right)\right]
\end{equation}
with the $\overline{ms}$ pole
\begin{equation}
\lambda=\frac{2}{4-d}+\ln 4\pi +\Gamma^\prime (1)+1.
\end{equation}\\

For $|\mu_\tin{I}|>m$ there is a symmetry breaking. The vacuum
expectation value that minimizes the potential, calculated in
,  is
\begin{eqnarray}
\bar U
 &=&
  \fr{m^2}{\mu_\tin{I}^2}
  +i[\tau_1\cos\phi+\tau_2\sin\phi]
   \sqrt{1-\fr{m^4}{\mu_\tin{I}^ 4}}\nonumber\\
 &\equiv&
  c+i\tilde\tau_1 s,
\end{eqnarray}
where $c\equiv \frac{m^2}{\mu_\tin{I}^2}$,
$s\equiv\sqrt{1-\fr{m^4}{\mu_\tin{I}^ 4}}$. From now, we will
refer with a tilde to any vector rotated in a $\phi$ angle
\begin{eqnarray}
\tilde v_1 &=& v_1\cos\phi+v_2\sin\phi\nonumber\\
\tilde v_2 &=& -v_1\sin\phi+v_2\cos\phi.
\end{eqnarray}

The expanded Lagrangian, keeping all the details, is given by
\begin{widetext}
\begin{eqnarray}
{\cal L}_{2,2}
 &=&
  \fr{1}{2}(\partial\bm{\pi})^2
   -\fr{1}{2}\mu_\tin{I}^2(s^2\tilde\pi_{\tin{1}}^2+\pi_{\tin{3}}^2)
  -|\mu_\tin{I}|c(\tilde\pi_{\tin{1}}\partial_{\tin{0}}\tilde\pi_{\tin{2}}
   -\tilde\pi_{\tin{2}}\partial_{\tin{0}}\tilde\pi_{\tin{1}})\nonumber\\
{\cal L}_{2,3}
 &=&
  \frac{1}{f}|\mu_\tin{I}|s\big[(\tilde\pi_{\tin{1}}^2+\pi_{\tin{3}}^2)
   \partial_{\tin{0}}\tilde\pi_{\tin{2}}
   -\fr{1}{2}|\mu_\tin{I}|c\tilde\pi_{\tin{1}}\bm{\pi}^2\big]\nonumber\\
{\cal L}_{2,4}
 &=&
  \frac{1}{f^2}\frac{1}{6}\Big\{\big[-(\partial\bm{\pi})^2
   +\mu_\tin{I}^2(s^2\tilde\pi_{\tin{1}}^2
    +\pi_{\tin{3}}^2
    -\fr{3}{4}c^2\bm{\pi}^2)
   +2|\mu_\tin{I}|c(\tilde\pi_{\tin{1}}\partial_{\tin{0}}\tilde\pi_{\tin{2}}
    -\tilde\pi_{\tin{2}}\partial_{\tin{0}}\tilde\pi_{\tin{1}})\big] \bm{\pi}^2
   +(\bm{\pi}\cdot\partial\bm{\pi})^2\Big\}\nonumber\\
{\cal L}_{4,1}
 &=&
  \frac{1}{f}\mu_\tin{I}^4cs[4s^2(l_1+l_2)-2c^2l_3-s^2l_4]\tilde\pi_{\tin{1}}\nonumber\\
{\cal L}_{4,2}
 &=&
  \frac{1}{f^2}\mu_\tin{I}^2\Big\{2s^2\big[(\partial\bm{\pi})^2
    +2(\partial_{\tin{0}}\tilde\pi_{\tin{2}})^2\big]l_1
    +2s^2\big[(\partial_{\tin{0}}\bm{\pi})^2
  +(\partial\tilde\pi_{\tin{2}})^2
    +(\partial_{\tin{0}}\tilde\pi_{\tin{2}})^2\big]l_2
    +\mu_\tin{I}^2c^2\epsilon_{\tin{ud}}^2\pi_3^2l_7\nonumber\\
  && \quad -2s^2\big[8|\mu_\tin{I}|c\tilde\pi_{\tin{1}}\partial_0\tilde\pi_{\tin{2}}
  +\mu_\tin{I}^2(s^2\bm{\pi}^2
     -3c^2\tilde\pi_{\tin{1}}^2
     -\tilde\pi_{\tin{2}}^2)\big](l_1+l_2)
  +\mu_\tin{I}^2c^2\big[s^2\tilde\pi_{\tin{1}}^2-c^2\bm{\pi}^2\big]l_3\nonumber\\
   && \quad +\big[c^2(\partial\bm{\pi})^2
        +2|\mu_\tin{I}|c(1-3c^2)
     \tilde\pi_{\tin{1}}\partial_{\tin{0}}\tilde\pi_{\tin{2}}
    -\mu_\tin{I}^2c^2\{(s^2+1)\bm{\pi}^2
      -(c^2-s^2)\tilde\pi_{\tin{1}}^2
      -\tilde\pi_{\tin{2}}^2\}\big]l_4\Big\},
\end{eqnarray}
\end{widetext}
where the index $n,m$ refers to ${\cal O}(p^n),{\cal O}(\pi^n)$
and $\epsilon_{\tin{ud}}^2 = (m_u-m_d)^2/(m_u+m_d)^2$. Note that
we use a negative chemical potential
($\mu_\tin{I}=-|\mu_\tin{I}|$) as it is the case in neutron stars
where there exist a condensate of $\pi_-$ cuasiparticles. The same
happens for the $\pi_+$ changing the sign of $\mu_\tin{I}$.

 From this
lagrangian we are able to extract the different vertices that will
participate in the diagrams responsible for mass renormalization.
Note that in the second phase, there will appear vertices with
three legs from (${\cal L}_{2,3}$) and one leg from (${\cal
L}_{4,1}$). The last one is responsible for the counterterms of
the tadpoles, but in the approximation we will use, these will not
be considered, since tadpoles, being of higher orders in the
expansion parameter, will be absent as we will see.

\section{Self-energy  in two different limits:
$|\mu_\tin{I}|\gtrsim m$, and $|\mu_\tin{I}|\gg m$}

In $\chi$PT the natural scale parameter is $\Lambda_\chi = 4\pi
f$, where $f$ is a parameter which at tree level coincides with
the pion decay constant. Usually $\Lambda_\chi$ is compared with
the tree level pion mass defining in this way the smallness
parameter for perturbative expansions ($\alpha = (\frac{m}{4\pi
f})^2$). Now, since $|\mu_\tin{I}|>m$ in the second phase, we will
re-define the smallness parameter as
\begin{equation}
\nu = \left(\frac{\mu_\tin{I}}{4\pi f}\right)^2
\end{equation}
This is possible for values of the chemical potential less than
$m_\eta$. For higher values in energy parameters we need to
consider the $SU(3)$ case.

The inverse of the free propagator for pions in momentum space,
i.e. the equations of motion extracted from ${\cal L}_{2,2}$ is
given by the matrix
\begin{equation}
  i\bm{D}^{-1}=\left(
   \begin{array}{ccc}
     p^2-\mu_\tin{I}^2s^2 & 2i|\mu_\tin{I}|cp_0 & 0          \\
     -2i|\mu_\tin{I}|cp_0 & p^2           & 0          \\
     0              & 0             & p^2-\mu_\tin{I}^2
   \end{array}
  \right),
\end{equation}
where
\begin{equation}
 \bm{D}_{ij}^{-1}=i\int d^4xe^{ipx}\frac{\delta^2
 S[\pi]}{\delta \pi_{\tin{i}}\delta\pi_{\tin{j}}}(x).
\end{equation}

The masses are defined as the poles of the determinant of the
propagator matrix at zero 3-momentum, i.e. the solution of
\begin{equation}
 \left|\bm{D^{-1}}(p)\right| _{\vec p=0}=0
\label{det}
\end{equation}

We can identify the $\pi_3=\pi_0$ field, as it is indicated in
\cite{Kogut:2001id}, through  its mass at $|\mu_\tin{I}|=m$ and because it is
diagonal in the propagator matrix. Therefore, there will be no
difficulties to handle this  propagator. This is not the case for
the charged pions which are mixed in a non trivial way.

The free propagator for the charged pions in momentum space is
\begin{equation}
  \tilde{\bm{D}}
    =\frac{i}{[p^4-p^2\mu_\tin{I}^2s^2-4\mu_\tin{I}^2c^2p_0^2]}
\left(
   \begin{array}{cc}
      p^2            & -2i|\mu_\tin{I}|cp_0 \\
      2i|\mu_\tin{I}|cp_0  & p^2-\mu_\tin{I}^2s^2\\
   \end{array}\right)
\end{equation}
with  $\tilde{\bm{D}}_{ij}\equiv\bm{D}_{ij\neq 3}$. The
Dolan-Jackiw propagators can be constructed with the general
formula
\begin{eqnarray}
\bm{D}(p;T) &=& \int \frac{dk_0}{2\pi i}\lim_{\eta\to 0}
\frac{\big[\bm{D}(k_0+i\eta,\bm{p})
    -\bm{D}(k_0-i\eta,\bm{p})\big]}{k_0-p_0-i\epsilon}\nonumber\\
&& \quad +n_B(p_0)\big[\bm{D}(p_0+i\epsilon,\bm{p})
  -\bm{D}(p_0-i\epsilon,\bm{p})\big].\nonumber\\
\end{eqnarray}

 The main
difficulty with this matrix propagator is that it is very
cumbersome to integrate the different loop corrections. This fact
motivated us to proceed in a systematic way, through an expansion
in a new appropriate smallness parameter, namely $s$ when
$|\mu_\tin{I}|\gtrsim m$  and $c$ when $|\mu_\tin{I}|\gg m$. We
realized that a similar expansion was proposed earlier by
Splittorff, Toublan and Verbaarschot 
\cite{Splittorff:2001fy,Splittorff:2002xn}.

It is not difficult to realize that the different vertices which
appear in our lagrangian will correspond to different  powers of
$c$ or $s$ depending on the case. Although it could appear as a
trivial correction, we will keep only the zero order in  our
calculations. As we will see, this procedure is not trivial at all
and provides us with interesting information about the behavior of
the pion masses as function of temperature and isospin chemical
potential.

If we scale all the parameters with $|\mu_\tin{I}|$ in all
structures, we have that:\\

{\it Propagators}
\begin{equation}
D(p;T,\mu_\tin{I};m)=\frac{1}{\mu_\tin{I}^2}\bar{\bm{D}}(\bar
p;\bar T,c)
\end{equation}\\

${\cal L}_4$ {\it constants}
\begin{equation}
l_i(\Lambda)=\frac{\gamma_i}{32\pi^2}\left[\bar l_i-\lambda
  -\ln (c\bar\Lambda^2)\right]
\end{equation}\\

{\it Vertices}
\begin{equation}
V_{n,m}(p;\mu_\tin{I};f,m,\Lambda)
    =\frac{|\mu_\tin{I}|^n}{f^{m+n-4}}v_{n,m}(\bar p;\bar\Lambda,c)
\end{equation}\\

{\it Integrals}
\begin{equation}
\int \frac{d^dk}{(2\pi)^d}\Lambda^{4-d}
    =\mu_\tin{I}^4\int\frac{d^d\bar k}{(2\pi)^d}\bar\Lambda^{4-d}
\end{equation}\\

{\it Self-energy}
\begin{equation}
\Sigma (p_0;T,\mu_\tin{I};f,m,\Lambda)
=\mu_\tin{I}^2\nu\sigma(\bar p_0;\bar T,c;\bar\Lambda)
\end{equation}\\
 where
from now on the bar on a parameter means that it is scaled with
$|\mu_\tin{I}|$.  It is possible then to expand the propagator in
powers of $s^n$ for$|\mu_\tin{I}|\gtrsim m$  and $c^n$ when
$|\mu_\tin{I}|\gg m$.

After the renormalization procedure, the divergent terms and the
scale factor $\lambda +\ln\bar\Lambda$ that will appear in the
loops calculation due to dimentional regularization and the ${\cal
L}_4$ terms will cancel. Finally the renormalized self-energy will
get the form
\begin{equation}
\begin{split}
  \Sigma_R\big|_ {|\mu_\tin{I}|\gtrsim m} =&\quad
   \mu_\tin{I}^2\nu\sum_{n=0}\sigma_n^{(s)}(\bar p;\bar T)s^n\\
  \Sigma_R\big| _{|\mu_\tin{I}|\gg m} =&\quad
   \mu_\tin{I}^2\nu\sum_{n=0}\big[\sigma_n^{(c)}(\bar p;\bar T)
    +\sigma^{\mbox{\tiny log}} _n(\bar p)\ln c \big]c^n.
   \end{split}
\end{equation}

As we said before we will keep only the $n=0$ terms in the
previous expansions. This approximation is certainly non trivial
since it allows us to explore the behavior of the renormalized
masses precisely in the vicinity of the transition point and also
in the region of the high chemical potential values. Further
corrections could also be calculated, taking into account higher
order vertices and propagators. We will consider these in a
further discussion.

Obviously, $c$ and $s$ must be smaller than the intersection point
$\frac{1}{\sqrt{2}}$. The idea is to find a region around the
intersection point that can be excluded safely from both
expansions. This excluded area becomes smaller when $n$ (the order
of the expansion) starts to grow. By demanding that
$c^{n+1},s^{n+1}\lesssim \frac{1}{2}(\frac{1}{\sqrt{2}})^{n+1}$,
we achieve this condition. Note that if $n=0$, we would exclude
the region of the chemical potential where $\sqrt{\frac{8}{7}}m^2
\lesssim \mui^2 \lesssim \sqrt{8}m^2$. We remark again that by
going to higher orders in our expansion, the excluded region
becomes smaller, so this is the best bound.

Note that bare masses (tree level masses) in this phase are
functions of the chemical potential as was discussed in
\cite{Kogut:2001id}
\begin{eqnarray}
m_0&=& |\mu_\tin{I}|\nonumber\\
m_+&=& |\mu_\tin{I}|\sqrt{1+3c^2}\nonumber\\
m_-&=& 0.
\end{eqnarray}

As the neutral pion propagator is diagonal with respect to the
charged ones, it's propagator will be the same in both limits at
any order:
\begin{equation}
D_{\tin{00}}(p;T) =
  \frac{i}{p^2-\mu_\tin{I}^2+i\epsilon}
  +2\pi n_B(|p_0|)\delta (p^2-\mu_\tin{I}^2).
\label{D00}
\end{equation}

\section{Renormalization and masses}\label{ram}

The corrected propagator  is
\begin{eqnarray}
\bm{D}_{c} &=& \bm{D}-i\bm{D}\bm{\Sigma}\bm{D}+\cdots\nonumber \\
  &=& \bm{D}(1+i\bm{\Sigma}\bm{D})^{-1}\nonumber\\
\bm{D}_{c}^{-1} &=& \bm{D}^{-1}+i\bm{\Sigma}.
\end{eqnarray}

Following the usual renormalization procedure, i.e. re-scaling the
fields $\pi_i=\sqrt{Z_i}\pi_i^r$ with $Z_i=1+{\cal O}(\nu)$ we
have that
\begin{eqnarray}
i D_{R}^{-1}(p)_\tin{ij}
  &=& Z_{ij}iD^{-1}(p)_\tin{ij}
  -\Sigma(p)_\tin{ij}+{\cal O}(\nu^2)\nonumber\\
  &=& iD^{-1}(p)_\tin{ij}
  -\Sigma_{R}(p)_\tin{ij}
\end{eqnarray}
with $ Z_{ij}\equiv\sqrt{Z_iZ_j}$. The value of $Z$ is chosen in
such a way that the corrected propagator does not have corrections
proportional to $p^2$.

As an example, let consider a free propagator and the self energy
correction of the form
\begin{eqnarray}
iD^{-1}(p) &=& p^2-2bp_0-a^2\nonumber\\
\Sigma (p) &=& \Sigma^\tin{0}+\Sigma^\tin{1}p_0+\Sigma^\tin{2}p^2.
\label{SigmaGeneral}
\end{eqnarray}
This expression for $\Sigma$ is valid in the first phase and we
will see that it is also valid in the second phase, up to  ${\cal
O}(s^0)$.

 Choosing $Z=1+\Sigma^2$ we have that the renormalized
self-energy will be
\begin{eqnarray}
\Sigma_R(p) &=& \big(\Sigma^\tin{0}+a^2\Sigma^\tin{2}\big)
 +\big(\Sigma^\tin{1}+2b\Sigma^\tin{2}\big)p_0\nonumber\\
  &=& \Sigma_R^\tin{0}+\Sigma_R^\tin{1}p_0.
\end{eqnarray}

Unfortunately, in general, the self-energy $\Sigma$  will not have
the shape of eq.(\ref{SigmaGeneral}), but it will be a complicated
function of the external momenta. As we want to compute mass
corrections, it is possible to expand the self energy in terms of
these corrections. In the rest frame, where $\bm{p}=0$, the energy
will be $p_0=m_R=m_t+\delta m$, where $m_R$ is the renormalized
mass, $m_t$ is the tree level mass and $\delta m$ is the
correction due to the self-energy terms. Then
\begin{eqnarray}
\Sigma (m_R) &=& \Sigma (m_t)
    +\Sigma ^\prime(m_t)\delta m
    +\fr{1}{2}\Sigma ^{\prime\prime}(m_t)(\delta m)^2
    +\dots\nonumber\\
    &=& \Sigma^\tin{0}[m_t]+\Sigma^{1}[m_t]m_R
    +\Sigma^\tin{2}[m_t]m_R^2+{\cal O}(\delta m)^3\nonumber\\
\end{eqnarray}
where the mass inside the brackets is the mass around which the
self-energy expansion was computed.

As we said  in eq.(\ref{det}), the masses are defined as the poles
of the determinant of the corrected propagator matrix at
$\bm{p}=0$. In the case of the neutral pion, since it does not mix
with the other terms, it is very easy to find the mass correction
as the solution of $D_{R}^{-1}(p_0)_\tin{00}=0$.

As we said before, due to the complicated form of the self-energy,
we expand the renormalized mass in the rest frame in powers of the
tree-level mass corrections $\delta m$. Since the renormalized
massed will be extracted as solutions of
\begin{equation}
\left| \tilde{\bm{ D}}^{-1}_R(m_R)\right|\equiv 0,
\end{equation}
then we only need to compute the $\delta m$ corrections. This can
be done taking an expansion of the determinant up to ${\cal
O}(\delta m)^2$, which leaves us with a quadratic equation for
$\delta m$.

After finding the $\delta m$ corrections, which are of course
given in a power series of $\nu$,  we can neglect terms of ${\cal
O}(\nu^2)$ in the renormalized mass. Note that for finite tree
level masses ($m_t \gg \nu|\mui|$) it is enough to expand up to
${\cal O}(\delta m)\sim {\cal O}(\nu)$

In the case of the condensed pion, where the tree-level mass
vanishes, some care has to be taken because now we do not have a
natural dimensionfull quantity to refer to as a scale parameter.
Here we will find corrections to the mass, only for finite
temperature, in the same way as it occurs for the photon-mass
corrections in QED 
\cite{LeBellac:1996}, which in our case will
include terms of the form $\sqrt{{\cal O}(\mui^2\nu)+{\cal
O}(\mui^2\nu^2)}$.


\section{Limit $|\mu_\tin{I}|\gtrsim m$}


\subsection{Propagators and vertices ($|\mu_\tin{I}|\gtrsim m$)}

\begin{figure} \vspace{1cm}
\fbox{
\includegraphics{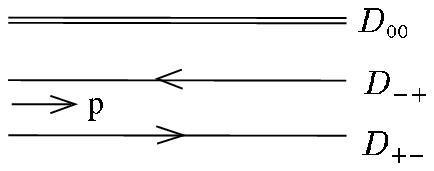}}
\caption{Propagators ${\cal O}(s^0)$ } \label{figprops}
\end{figure}

\begin{figure}
\vspace{1cm} \fbox{
\includegraphics{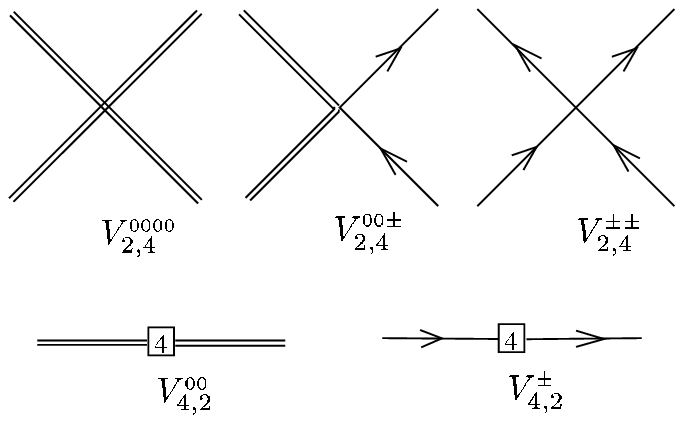}}
\caption{Relevant vertices at $ {\cal O}(s^0)$} \label{verts}
\end{figure}

The propagator for the charged pions at ${\cal O}(s^0)$ is given
by
\begin{equation}
  \tilde{\bm{D}}
    =\frac{i}{[p^4-4\mu_\tin{I}^2p_0^2]}
\left(
   \begin{array}{cc}
      p^2             & -2i|\mu_\tin{I}|p_0 \\
      2i|\mu_\tin{I}|p_0  & p^2\\
   \end{array}\right)+\frac{1}{\mu_\tin{I}^2}{\cal O}(s^2)
\end{equation}
It is more convenient to work with the following combination of
fields
\begin{equation}
\tilde\pi_{\pm}  =\fr{1}{\sqrt{2}}(\tilde\pi_{\tin{1}}\mp
i\tilde\pi_{\tin{2}}).
\end{equation}
We remark that the $\tilde\pi_{\pm}$ fields do not correspond to
the physical charged pion fields but to a combination of them.
This fact is a consequence of the non trivial vacuum structure in
this phase. This combination is also not trivial due to derivative
terms and the inverse D'alambertian operator. We would like to
mention that in the first phase ($\mui < m$) the $\pi_\pm$
correspond effectively to the charged pions and the propagator
becomes anti-diagonal.
 Then, the equations of motion for charged pions are
\begin{eqnarray}
 i\tilde{\bm{D}}^{-1}\!\!\!\!
  &=& \!\!\!\!\left(
   \begin{array}{cc}
    -\fr{1}{2}\mu_\tin{I}^2s^2
    & p^2+2|\mu_\tin{I}|cp_0-\fr{1}{2}\mu_\tin{I}^2s^2 \\
    &                                       \\
    p^2-2|\mu_\tin{I}|cp_0-\fr{1}{2}\mu_\tin{I}^2s^2
    & -\fr{1}{2}\mu_\tin{I}^2s^2
   \end{array}\right)\nonumber\\
    && \nonumber\\
   &=& i\left(
  \begin{array}{cc}
   D^{-1}_\tin{++} &  D^{-1}_\tin{+-} \\
                   &                  \\
   D^{-1}_\tin{-+} &  D^{-1}_\tin{--}
  \end{array}\right)
\end{eqnarray}
and the free propagator matrix is
\begin{equation}
  \tilde{\bm{D}} = i\left(
   \begin{array}{cc}
      0                         & \frac{1}{p^2-2|\mu_\tin{I}|p_0} \\
      \frac{1}{p^2+2|\mu_\tin{I}|p_0}  &                         0\\
   \end{array}\right)+\frac{1}{\mu_\tin{I}^2}{\cal O}(s).
\end{equation}

The Dolan-Jackiw propagators for charged pions, including isospin
chemical potential (enough for one loop calculations) are
\begin{eqnarray}
  D(p;T)_{\tin{+-}} &=&
  \frac{i}{p^2-2|\mu_\tin{I}|p_0+i\epsilon}\nonumber\\
 && +2\pi n_B(|p_0|)\delta (p^2-2|\mu_\tin{I}|p_0)
   +\frac{1}{\mu_\tin{I}^2}{\cal O}(s^2)\nonumber\\
 D(p;T)_\tin{-+} &=& D(-p;T)_\tin{+-}
\end{eqnarray}
and the other propagators, $D_\tin{++}$ and $D_\tin{--}$  are of
order $s^2$, so we will neglect them. For diagrammatic purposes,
the $D_{\tin{+-}}$ propagator will be drawn as an arrow from $+$
to $-$ as appear in FIG. \ref{figprops}.

The relevant vertices at the zero order in this region are shown
in FIG. \ref{verts}, where the corresponding analytical expressions are given
by
\begin{widetext}
\begin{eqnarray}
 V_{2,4}^\tin{0000} &=&
  i\frac{\mui^2}{f^2}\big[1+{\cal O}(s^2)\big] \nonumber\\
V_{2,4}^\tin{00\pm} &=&
 \frac{i}{3f^2}\Big\{ 2p_{\tin{(0)}}q_{\tin{(0)}}
   +2p_{\tin{(+)}}p_{\tin{(-)}}
   +|\mui|(p_{\tin{(+)}}^0-p_{\tin{(-)}}^0)-\mui^2
   -\big[p_{\tin{(0)}}+q_{\tin{(0)}}\big]
   \big[p_{\tin{(+)}}+p_{\tin{(-)}}\big]
  +\mu_\tin{I}^2{\cal O}(s^2)\Big\}\nonumber\\
V_{2,4}^{\tin{\pm\pm}} &=&
 \frac{i}{3f^2}\Big\{ \big[p_{\tin{(+)}}+q_{\tin{(+)}}\big]
  \big[p_{\tin{(-)}}+q_{\tin{(-)}}\big]
  -6\mu_\tin{I}^2 
    -4|\mui| (p_{\tin{(+)}}^0+q_{\tin{(+)}}^0
  -p_{\tin{(-)}}^0-q_{\tin{(-)}}^0) -2p_{\tin{(+)}}q_{\tin{(+)}}
    -2p_{\tin{(-)}}q_{\tin{(-)}}
  +\mui^2{\cal O}(s^2)\Big\} \nonumber\\
V_{4,2}^{\tin{00}} &=&
    -i2\frac{\mui^2}{f^2}
        \big[\mui^2l_3+(2pq
        +\mui^2)l_4-\mui^2\epsilon_{ud}^2l_7+\mui^2{\cal O}(s^2)\big]\nonumber
\\
V_{4,2}^{\tin{-+}} &=&
    -i2\frac{\mui^2}{f^2}\big[\mui^2l_3+(p_{\tin{(+)}}p_{\tin{(-)}}
    +|\mui| (p_{\tin{(+)}}^0-p_{\tin{(-)}}^0))l_4+\mu_\tin{I}^2{\cal
    O}(s^2)\big].
\end{eqnarray}
\end{widetext}
 All the momenta emerge from the
vertices. In principle, there will appear also other vertices:
$v_{2,2}^{\tin{++}}$, $v_{2,2}^{\tin{--}}$,
$v_{2,4}^{\tin{00++}}$, $v_{2,4}^{\tin{00--}}$, $v_{2,4}^{\tin{\pm
++}}$, $v_{2,4}^{\tin{\pm --}}$  of ${\cal O}(s^2)$ and
 $v_{2,3}$, $v_{4,1}$  of ${\cal O}(s)$. As we said, however, since we are working at
 zero order, this terms will not be considered in our analysis.
 
\subsection{Self-energy ($|\mu_\tin{I}|\gtrsim m$)}

\begin{figure}
\fbox{
\includegraphics{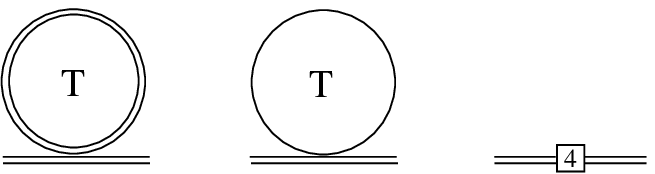}}
\caption{$\Sigma_{\tin{00}}$ at  ${\cal O}(s^0)$} \label{S00s}
\end{figure}

\begin{figure}
\fbox{
\includegraphics{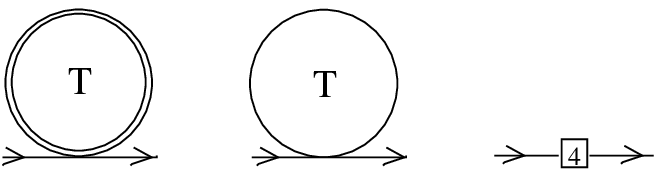}}
\caption{$\Sigma_{\tin{-+}}$ at  ${\cal O}(s^0)$} \label{Smps}
\end{figure}

Proceeding with the relevant vertices and propagators the loops
corrections to the  pion's propagator matrix are shown in FIG.
\ref{S00s} and FIG. \ref{Smps}

Defining $\underline\lambda = \lambda +\ln\bar\Lambda$, the
self-energy is
\begin{widetext}
\begin{eqnarray}
 \Sigma(p)_{\tin{00}}
  &=&\nu\Big\{-\mu_\tin{I}^2\big[ \fr{4}{3}\underline\lambda
   +\fr{1}{2}\bar l_3 -2\bar l_4
   +2I'_0-\fr{4}{3}I'\big] +p^2\big[\fr{4}{3}\underline\lambda
   -2\bar l_4+\fr{8}{3}I'\big]+{\cal O}(\mui^2s^2)\Big\}\nonumber\\
 \Sigma(p)_{\tin{-+}}
  &=& \nu\Big\{ -\mu_\tin{I}^2\big[\fr{1}{2}\bar l_3-2I'_0-8J'\big]-2p_0|\mu_\tin{I}|\big[\fr{4}{3}\underline\lambda-2\bar l_4
    +\fr{4}{3}I'_0+\fr{4}{3}I'+4J'\big] +p^2\big[\fr{4}{3}\underline\lambda
    -2\bar l_4+\fr{4}{3}I'_0+\fr{4}{3}I'\big]
    +{\cal O}(\mui^2 s^2)\Big\}\nonumber\\
 \Sigma(p)_{\tin{+-}}
  &=& \Sigma(-p)_{\tin{-+}}
\end{eqnarray}
\end{widetext}
with
\begin{eqnarray}
I' &\equiv& \int_1^\infty dx\sqrt{x^2-1}
 \Big[ n_B\big(|\mui ||x-1|\big)
  +x\rightarrow -x\Big]\nonumber\\
J'&\equiv& \int_1^\infty dx\sqrt{x^2-1}
 \Big[ xn_B\big(|\mui ||x-1|\big)
  + x\rightarrow -x\Big]\nonumber\\
I'_n &\equiv&
 \int_1^\infty dx\sqrt{x^2-1}x^{2n}2n_B\big(|\mu_\tin{I}|x\big).
\end{eqnarray}
Note that these definitions are almost the same ones we used in
our previous paper 
\cite{Loewe:2002tw,Loewe:2003dq}. Here, however, the term that
multiplies $x$ in the argument of the Bose-Einstein distribution
is $\mui $ instead of $m$.
Following the prescription indicated in Section III, that the
renormalized self-energy does not depend on $p^2$ we have that the
renormalization constants are
\begin{eqnarray}
 Z_0 &=&  1+\nu \big[\fr{4}{3}\underline\lambda
   -2\bar l_4+\fr{8}{3}I'\big]\nonumber\\
 Z_\pm &=& 1+\nu\big[ \fr{4}{3}\underline\lambda
   -2\bar l_4 +\frac{4}{3}I'_0+\fr{4}{3}I'\big]
\end{eqnarray}
and the renormalized self-energy corrections are
\begin{eqnarray}
\Sigma_{R\tin{00}} &=& \mu_\tin{I}^2\nu
 \big[-\fr{1}{2}\bar l_3-2I'_0+4I'\big]\nonumber\\
\Sigma_{R}(p_0)_\tin{-+} &=& \nu\Big\{\mu_\tin{I}^2
 \big[-\fr{1}{2}\bar l_3+2I'_0+8J'\big]
    -8|\mu_\tin{I}|J'p_0\Big\}\nonumber\\
\Sigma_{R}(p_0)_\tin{+-}
  &=& \Sigma_{R} (-p_0)_\tin{-+}
\end{eqnarray}
plus higher corrections of order $\mui^2\nu s^2$. We will use
these results to extract the renormalized temperature- and
chemical potential-dependent masses.

\subsection{Masses ($|\mu_\tin{I}|\gtrsim m$)}

To extract the  $m_{\pi^\tin{0}}$ mass, we do not need any further
effort  since it is already diagonal in the matrix propagator, and
it's renormalized self-energy is constant in the external momenta,
i.e., it does not depend on $p_0$.

\begin{equation}
m_{\pi^\tin{0}}=|\mui|\Big\{ 1+\nu\big[ -\fr{1}{4}\bar
l_3-I'_0+2I'\big]\Big\}.
\end{equation}

For the case of $m_{\pi^\tin{\pm}}$, the vanishing determinant of
the charged matrix propagator, provides us with a second order
equation in powers of $p_0^2$.
\begin{eqnarray}
&& p_0^4-p_0^2\Big[ m_+^2 +2\Sigma_{R\tin{-+}}^\tin{0}
+4c\Sigma_{R\tin{-+}}^\tin{1}
+(\Sigma_{R\tin{-+}}^\tin{1})^2\Big]\nonumber\\
&& \qquad +s^2\Sigma_{R\tin{-+}}^\tin{0}
+(\Sigma_{R\tin{-+}}^\tin{0})^2=0.
\end{eqnarray}
The solution of this equation gives the $\pi^\pm$ masses,
recognizing them respectively according to the tree-level case if
we set the self-energy corrections to zero. We can write the
expression for the masses in a more compact way, by expanding
around the $m_+$ term that appears in the last equation. Then, the
masses are given by
\begin{eqnarray}
m_{\pi^+} &=& m_+ +\nu |\mui|\big[ -\fr{1}{4}\bar l_3 +I'_0-4J' +{\cal O}(s^2)\big]\nonumber\\
m_{\pi^-} &=&\fr{1}{2}|\mui|\Big\{\big( s^2 +\nu[ -\fr{1}{2}\bar
l_3 +2I'_0 +8J']\big)\nonumber\\
    && \quad \times  \nu [ -\fr{1}{2}\bar l_3 +2I'_0
+8J']  +{\cal O}(\nu^2s^2)\Big\}^{1/2}.
\end{eqnarray}

As we indicated previously,  $m_{\pi^\tin{-}}$ can not be
expressed as a sum of some initial mass plus corrections, since it
does not have a tree level mass to compare with in the second
phase. Thermal radiative corrections, however, will induce a non
trivial behavior for the renormalized $m_{\pi^-}(T,\mui)$. It is
interesting to remark that $m_{\pi^\tin{-}}$ does not have a
finite real solution for $\bar l_3 > 4I'_0+16J'$, according to the
previous equation. This means that $\pi^-$ remains in the
condensed state for $T$ less than a certain critical value.

\section{Limit $|\mu_I|\gg m$}

\subsection{Propagators and vertices ($|\mu_I|\gg m$)}

\begin{figure}
\fbox{
\includegraphics{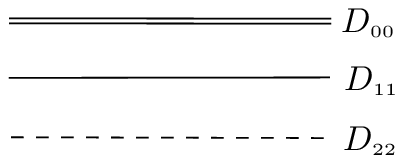}}
\caption{Propagators ${\cal O}(c^0)$ }
 \label{propc}
\end{figure}

\begin{figure}
\fbox{
\includegraphics{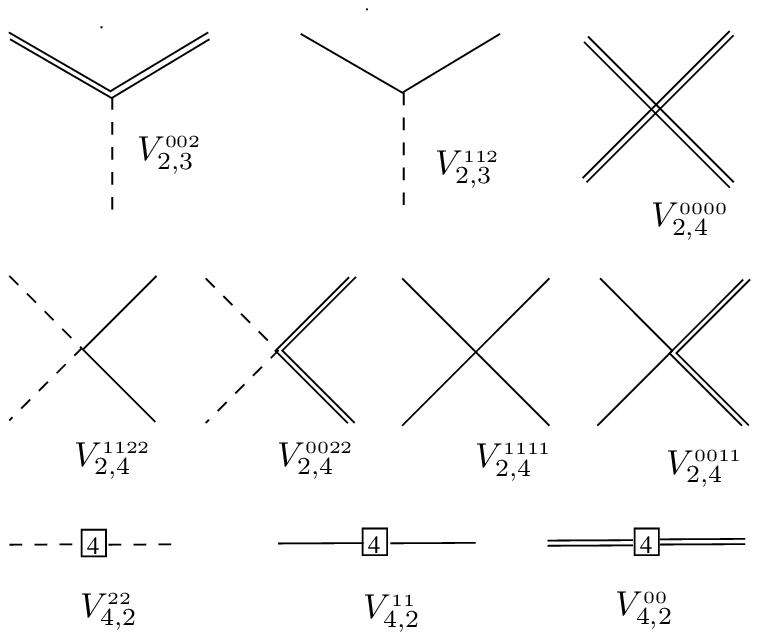}}
\caption{Relevant vertices at $ {\cal O}(c^0)$} \label{vertc}
\end{figure}

Proceeding in the same way as in the $|\mu_I|\gtrsim m$ case, the
propagator of the charged pions $\tilde\pi_{\tin{1}}$ and
$\tilde\pi_{\tin{2}}$ at zero temperature at ${\cal O}(c^0)$ is
\begin{equation}
 \tilde{\bm{D}}
    =i
\left(
   \begin{array}{cc}
      \frac{1}{p^2-\mu_I^2} & 0 \\
                     0      & \frac{1}{p^2} \\
   \end{array}\right)+\frac{1}{\mu_I^2}{\cal O}(c).
\end{equation}

The $\pi_{\tin{0}}$ propagator remains the same as the
$|\mu_I|\gtrsim m$ case in the equation (\ref{D00}). Then, the
Dolan-Jackiw propagators for charged pions are
\begin{equation}
\begin{split}
D_{\tin{11}}(p;T,|\mui|)=& D_{\tin{00}}(p;T,|\mui|)+\frac{1}{\mu_I^2}{\cal O}(c^2)\\
D_{\tin{22}}(p;T,|\mu_I|)=& \frac{i}{p^2+i\epsilon} +2\pi
n_B(|p^0|)\delta (p^2)+\frac{1}{\mu_I^2}{\cal O}(c^2)
\end{split}
\end{equation}
 and the propagators $\bar D_{\tin{12}}$ and
 $\bar D_{\tin{21}}$ are of order $c$. In the case of a
 propagator with zero mass, it is necessary to introduce a small
 fictitious mass as a regulator, i.e.
 $D_{\tin{22}}^{-1}=\lim_{\eta\to 0}(p^2-\eta^2)$.
Note that in  the chiral limit ($c=0$)  the fields $\pi_{\tin{1}}$
and $\pi_{\tin{0}}$ have the same behavior. The other components
of the propagator matrix are of order $c$.
Diagrammatically, we
will denote the $D_{\tin{11}}$ propagator with a line and
$D_{\tin{22}}$ with a dashed line, as we can see in FIG.
\ref{propc}. The $D_{\tin{00}}$ propagator remains the same.

The relevant vertices at ${\cal O}(c^0)$ are indicated in FIG.
\ref{vertc} with the analytical expressions
\begin{widetext}
 \begin{eqnarray}
 V_{2,3}^{\tin{002}} &=& V_{2,3}^{\tin{112}} =
   \frac{2|\mui|}{f}\big[ p_\tin{(2)}^0 +{\cal O}(\mui c)\big]\nonumber\\
 V_{2,4}^{\tin{0000}} &=&
  i\frac{\mu_I^2}{f^2}[1+{\cal O}(c^2)]\nonumber\\
 V_{2,4}^{\tin{0011}} &=&
  \frac{i}{3f^2}[2p_{\tin{(0)}}q_{\tin{(0)}}+2p_{\tin{(1)}}q_{\tin{(1)}}
    +4\mui^2
   -(p_{\tin{(0)}}+q_{\tin{(0)}})
    (p_{\tin{(1)}}+q_{\tin{(1)}})+\mui^2{\cal O}(c^2)]\nonumber\\
 V_{2,4}^{\tin{0022}} &=&
  \frac{i}{3f^2}[2p_{\tin{(0)}}q_{\tin{(0)}}+2p_{\tin{(2)}}q_{\tin{(2)}}
   +2\mui^2-(p_{\tin{(0)}}+q_{\tin{(0)}})
    (p_{\tin{(2)}}+q_{\tin{(2)}})+\mui^2{\cal O}(c^2)]\\
V_{2,4}^{\tin{1122}} &=&
  \frac{i}{3f^2}[2p_{\tin{(1)}}q_{\tin{(1)}}+2p_{\tin{(2)}}q_{\tin{(2)}}
    +2\mui^2 -(p_{\tin{(1)}}+q_{\tin{(1)}})
    (p_{\tin{(2)}}+q_{\tin{(2)}})+\mui^2{\cal O}(c^2)]\nonumber\\
 V_{4,2}^{\tin{00}} &=&
  i4\frac{\mui^2}{f^2}[(p_{\tin{(0)}}q_{\tin{(0)}}+\mui^2)l_1 
    +(p_{\tin{(0)}}^0q_{\tin{(0)}}^0+\mu_I^2)l_2+\mui^2{\cal O}(c^2)]\nonumber\\
 V_{4,2}^{\tin{11}} &=&
  i4\frac{\mui^2}{f^2}[(p_{\tin{(1)}}q_{\tin{(1)}}+\mui^2)l_1 
    +(p_{\tin{(1)}}^0q_{\tin{(1)}}^0+\mui^2)l_2+\mui^2{\cal O}(c^2)]\nonumber\\
 V_{4,2}^{\tin{22}} &=&
  i4\frac{\mui^2}{f^2}[(p_{\tin{(2)}}q_{\tin{(2)}}
   +2p_{\tin{(2)}}^0q_{\tin{(2)}}^0)(l_1+l_2)+\mui^2{\cal
   O}(c^2)].\nonumber
\end{eqnarray}
\end{widetext}
As in the $|\mui |\gtrsim m$ case, there will appear also other
vertices of higher order in power of $c$: $\bar
V_{2,2}^{\tin{$12$}}$, $v_{2,3}^{\tin{$001$}}$,
 $v_{2,3}^{\tin{$111$}}$, $v_{2,3}^{\tin{$122$}}$,
 $v_{2,4}^{\tin{$0012$}}$, $v_{2,4}^{\tin{$1112$}}$,
 $v_{2,4}^{\tin{$1222$}}$, $v_{4,2}^{\tin{$12$}}$, of ${\cal O}(c)$.

\subsection{Self-Energy ($|\mu_I|\gg m$)}

\begin{figure}
\vspace{1cm} \fbox{
\includegraphics{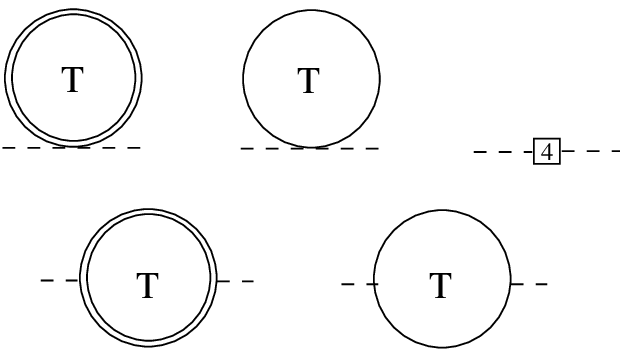}}
\caption{$\Sigma_{\tin{22}}$ at  ${\cal O}(c^0)$}
 \label{S22c}
\end{figure}

\begin{figure}
\vspace{1cm} \fbox{
\includegraphics{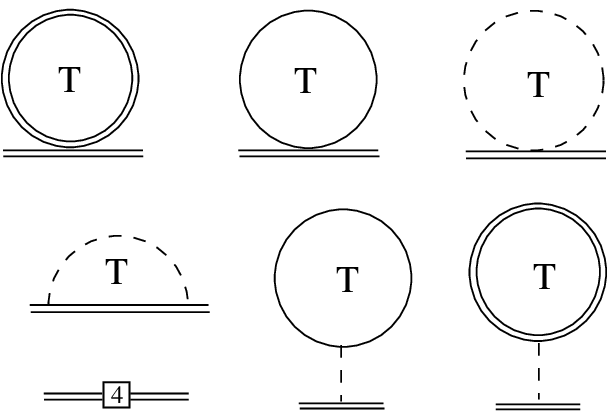}}
\caption{$\Sigma_{\tin{00}}$ (or $\Sigma_{\tin{11}}$) at  ${\cal
O}(c^0)$}
 \label{S00c}
\end{figure}

As we did for the $|\mui |\gtrsim m$ case, we use the relevant
vertices and propagators to compute the self-energy corrections.
In our approximation (${\cal O}(c^0)$), the corrections to the
$D_{\tin{00}}$ and $D_{\tin{11}}$ are the same. This happens
because their vertices and propagators differ in quantities of
order higher than $c$.

The loop corrections are shown in FIG. \ref{S00c}, for
$D_{\tin{11}}$ (which are the same as those of $D_{\tin{00}}$
exchanging the single line with the double line), and FIG.
\ref{S22c} for $D_{\tin{22}}$.

Note that in this region tadpole diagrams appear, which are absent
in the previous case. However, at the order $c^0$, it turns out
that these tadpoles vanish, because the vertex is proportional to
$p_0$, and the tail of the tadpole does not carry momentum. At the
order $c$, there will appear non vanishing tadpole diagrams (and
also new ${\cal L}_4$ corrections). Nevertheless, for high values
of the chemical potential, the leading behavior will be given by
our approximation.

The self energy corrections are
\begin{widetext}
\begin{eqnarray}
\Sigma(p_0)_\tin{00} &=& \nu\mui^2 \Big\{ (p_0^2-1)\big[
\fr{2}{3}\underline\lambda +2\ln c -\fr{2}{3}\bar l_1
-\fr{4}{3}\bar l_2 +\fr{4}{3}+\fr{4}{3}I'_0 +\fr{4}{9}\pi^2T^2\big]
       \nonumber\\
      && \hspace{1cm} +\fr{4}{3} -4I'_0+2A_1(p_0)+16B_0(p_0) +16B_1(p_0) -32\pi iC_1(p_0)\Big\}\nonumber\\
\Sigma(p_0)_\tin{11} &=& \Sigma(p_0)_\tin{00}\nonumber\\
 \Sigma(p_0)_\tin{22}
&=& 2\nu\mui^2\Big\{ p_0^2\big[ \fr{2}{3}\underline\lambda -\bar
l_1 -2\bar l_2 +3\ln c +2
-\fr{4}{3}I'_0\big]+2A_2(p_0) +16B_2(p_0) -32\pi iC_2(p_0)\Big\}
\end{eqnarray}
\end{widetext}
plus corrections of order $\nu\mui^2 c^2$. The functions $A_n$,
$B_n$, $C_n$ are defined as
\begin{widetext}
\begin{eqnarray}
A_1(p_0) &=& \int_0^1dx\big[ 3p_0^2x^2 +(p_0^2-1)x\big]
    \ln\big[p_0^2(x^2-x) +x -i\epsilon\big]\nonumber\\
A_2(p_0) &=& \int_0^1dxp_0^2\ln \big[ p_0^2(x^2-x) +1
-i\epsilon\big]\nonumber\\
B_0(p_0) &=& \int_0^\infty dxxn_B(|\mui|x)\left[\frac{x^2}{p_0^2
-2p_0x -1 +i\epsilon}+x\rightarrow
    -x\right]\nonumber\\
B_1(p_0) &=& \int_1^\infty
dx\sqrt{x^2-1}n_B(|\mui|x)\left[\frac{(p_0 -x)^2}{p_0^2-2p_0x +1
+i\epsilon}+x\rightarrow
    -x\right]\nonumber\\
B_2(p_0) &=& \int_1^\infty dx\sqrt{x^2-1}n_B(|\mui|x)
\left[\frac{p_0^2}{p_0^2-2p_0x+i\epsilon}+x\rightarrow
    -x\right]\nonumber\\
C_1(p_0) &=& \left| \frac{p_0^2-1}{2p_0}\right| n_B\!\!\left(
\left| \frac{p_0^2-1}{2p_0}\right|\right) n_B\!\!\left( \left|
\frac{p_0^2+1}{2p_0}\right|\right)\nonumber\\
C_2(p_0) &=& \big[ \theta (p_0-2)+\theta
(-p_0-2)\big]p_0^2\sqrt{(p_0/2)^2-1}n_B(|p_0/2|)^2.
\end{eqnarray}
\end{widetext}

As we can see from the previous equations, it is by far non
trivial to identify the renormalized mass, as it was the case in
the region $|\mui |\gtrsim m$. Therefore we need to expand the
propagator, as we explained in Section \ref{ram}, around the
tree-level masses, identifying the term proportional to $p_0^2$,
or in this case $(m_t+\delta m)^2$.

For the case of $\Sigma_{R\tin{00}}$, it is enough to evaluate
$\Sigma_\tin{00}$ with $p_0=m_{\pi^0}=m_0+\delta m_0$ because
$D_{R\tin{00}}^{-1}$ is diagonal with respect to the charged pions
propagator. Then we expand around $m_0$ and renormalize it,
according to what we explained in  Section \ref{ram}. For
$\Sigma_{R\tin{11}}$ and $\Sigma_{R\tin{11}}$ we need to evaluate
it in two values, $m_{\pi^{\pm}}=m_\pm +\delta m_\pm$.

The renormalization constants for the different masses are
(evaluated at these masses)
\begin{widetext}
\begin{eqnarray}
Z_{0}[m_0] &=& 1+2\nu\Big\{ \fr{1}{3}\big[ \underline\lambda +2\ln
c -\bar l_1 -2\bar l_2\big] -7 +\fr{38}{9}\pi^2\fr{T^2}{\mui^2}
   -\fr{34}{3}I'_0 -2K_{21}
    -16\pi i\fr{T}{|\mui|}n_B(|\mui|)\Big\}\nonumber\\
Z_1[m_+] &=& Z_0+{\cal O}(\nu c^2)\nonumber\\
Z_1[m_-] &=& 1+\fr{2}{3}\nu\Big\{ \underline\lambda +3\ln c -\bar
l_1 -2\bar l_2 +\fr{3}{4} +\fr{2}{3}\pi^2\fr{T^2}{\mui^2}
   -\fr{16}{5}\pi^4\fr{T^4}{\mui^4}
-\fr{512}{21}\pi^6\fr{T^6}{\mui^6} -\fr{34}{3}I'_0 -6K_{21}\Big\}\nonumber\\
Z_2[m_+] &=& 1+2\nu\Big\{ \fr{2}{3}\underline\lambda \bar l_1
-2\bar l_2 +3\ln c +\fr{4}{3} -\fr{10\pi}{9\sqrt{3}}
    -8K_{10} -2K_{12}
    +{\cal O}(c^2)\Big\}\nonumber\\
Z_2[m_-] &=&  1+2\nu\Big\{ \fr{2}{3}\underline\lambda \bar l_1
-2\bar l_2 +3\ln c +2 -\fr{4}{3}I'_0 -8K_{00}\Big\}\nonumber\\
\end{eqnarray}
\end{widetext}
where the mass inside the brackets is the mass around which the
self-energy expansion was computed. The function $K_{nm}$ is
defined as
\begin{equation}
K_{nm}\equiv \int_1^\infty dx\frac{\sqrt{x^2-1}n_B(|\mui|x)}{\big
[x^2-(n/2)^2\big]^{m+1}}.
\end{equation}

The self-energy corrections, associated to the different
tree-level pion masses, as function of the corresponding
renormalized masses are
\begin{widetext}
\begin{eqnarray}
\Sigma_R(m_{\pi^0})_\tin{00} &=& 4\nu|\mui|\Big\{ \big[ 6
-\fr{14}{3}\pi^2\fr{T^2}{\mui^2} +13I'_0 +2K_{21}
     +16\pi i\fr{T}{|\mui|}n_B(|\mui|)\big]
(m_{\pi^0}-|\mui|) +3I'_0\Big\}\nonumber\\
 \Sigma_R(m_{\pi^+})_\tin{11} &=&
\Sigma_R(m_{\pi^+})_\tin{00}
+{\cal O}(\mui^2\nu c^2)\nonumber\\
\Sigma_R(m_{\pi^-})_\tin{11}  &=& \!\!\! -2\nu\mui^2\Big\{
\fr{1}{6} +\fr{512}{63}\pi^6\fr{T^6}{\mui^6} +16I'_0 -8I'_1
-2K_{21}\Big\}\nonumber\\
 \Sigma_R(m_{\pi^+})_\tin{22} &=& 2\nu|\mui|\Big\{
\big[
-\fr{2}{3} -\fr{4\pi}{9\sqrt{3}} +16K_{10} -5K_{11}\big]|\mui|
    +\big[ -\fr{8}{3} +\fr{32\pi}{9\sqrt{3}} +20K_{11}
    +3K_{12}\big]m_{\pi^+}\Big\}
    +{\cal O}(\mui^2\nu c^2)\nonumber\\
\Sigma_R(m_{\pi^-})_\tin{22} &=& {\cal O}(\mui^2\nu c^2)
\end{eqnarray}
\end{widetext}
Note that the self-energy corrections actually have the form
$\Sigma_R(m_{\pi^i}) =
\Sigma_R^\tin{0}[m_i]+\Sigma_R^\tin{1}[m_i]m_{\pi^i}$
 presented in Section \ref{ram}

\subsection{Masses ($|\mu_I|\gg m$)}
Since we have already expanded the self-energy corrections in
powers  of the mass corrections $(\delta m_i)^n$, we can neglect
higher terms in the determinant of the propagator  finding then
the solution for $\delta m_i$ from the pole condition. For
$m_{\pi^0}$ and $m_{\pi^+}$ the corrections $\delta m_0$, $\delta
m_+$ are of order $\mui\nu$ (neglecting terms of order $\nu c$).
For $m_{\pi^-}$, because there is no finite tree-level mass, we
need to keep in the expansion terms of the order $\nu^2$; i.e. we
can consider, in principle, $(\delta m_-)^2 \sim \mui\nu^{1/2}$.
Remembering that $\Sigma_R^\tin{0}$ of order $\mui^2\nu$, and
$\Sigma_R^\tin{1}$ of order $\mui\nu$, we can expand the
propagator, neglecting higher order terms.

Summarizing the power counting is:
\begin{eqnarray}
\delta m_0 &\sim& {\cal O}(\mui\nu)\nonumber\\
\delta m_+ &\sim& {\cal O}(\mui\nu)\nonumber\\
\delta m_- &\sim& {\cal O}(\mui\nu^{1/2})\nonumber\\
\Sigma_R^\tin{0} &\sim& {\cal O}(\mui^2\nu)\nonumber\\
\Sigma_R^\tin{1} &\sim& {\cal O}(\mui\nu).
\end{eqnarray}
The resulting expressions for the renormalized  masses are
surprisingly simple. Many terms vanish, including also some
complex contributions.

\begin{eqnarray}
m_{\pi^0} &=& |\mui|\big[ 1+6\nu I'_0\big]\nonumber\\
m_{\pi^+} &=& |\mui|\big[ \sqrt{1+3c^2} +6\nu I'_0\big]\nonumber\\
m_{\pi^-} &=& {\cal O}(\mui\nu c^2).
\end{eqnarray}

Note that $m_{\pi^-}$ vanishes once again in this region, i.e.,
the pion condenses again, in spite of the fact that the thermal
corrected mass started to grow near the phase transition point.
This behavior, however,  could be just a fictitious result from
our expansion up to order $c^0$.

\section{Results}

\begin{figure}
\vspace{0cm}
 \fbox{
\includegraphics[scale=.6]{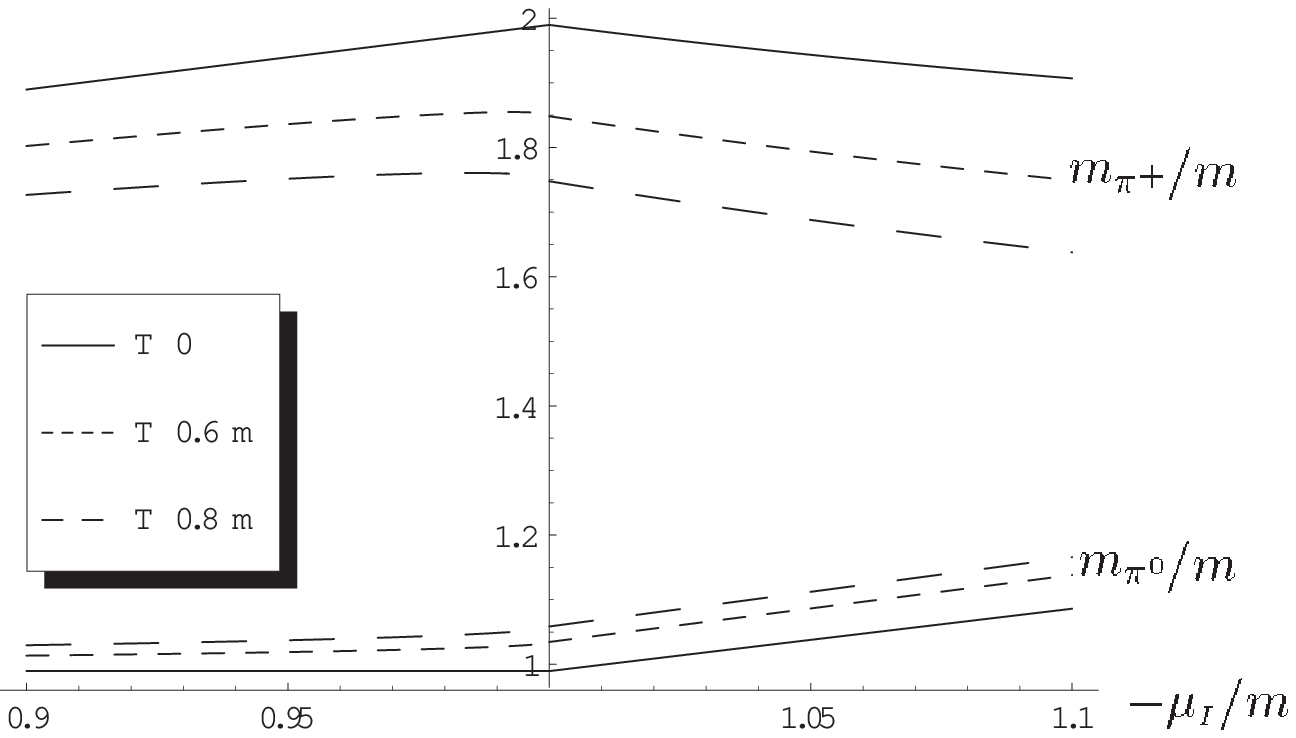}}
\caption{$m_{\pi^0}$ and $m_{\pi^+}$ as a function of the isospin
chemical potential at different values of the temperature. All
parameters are scaled with $m$.}
 \label{m0mps}
\end{figure}

\begin{figure}
\vspace{0cm}
 \fbox{
\includegraphics[scale=.55]{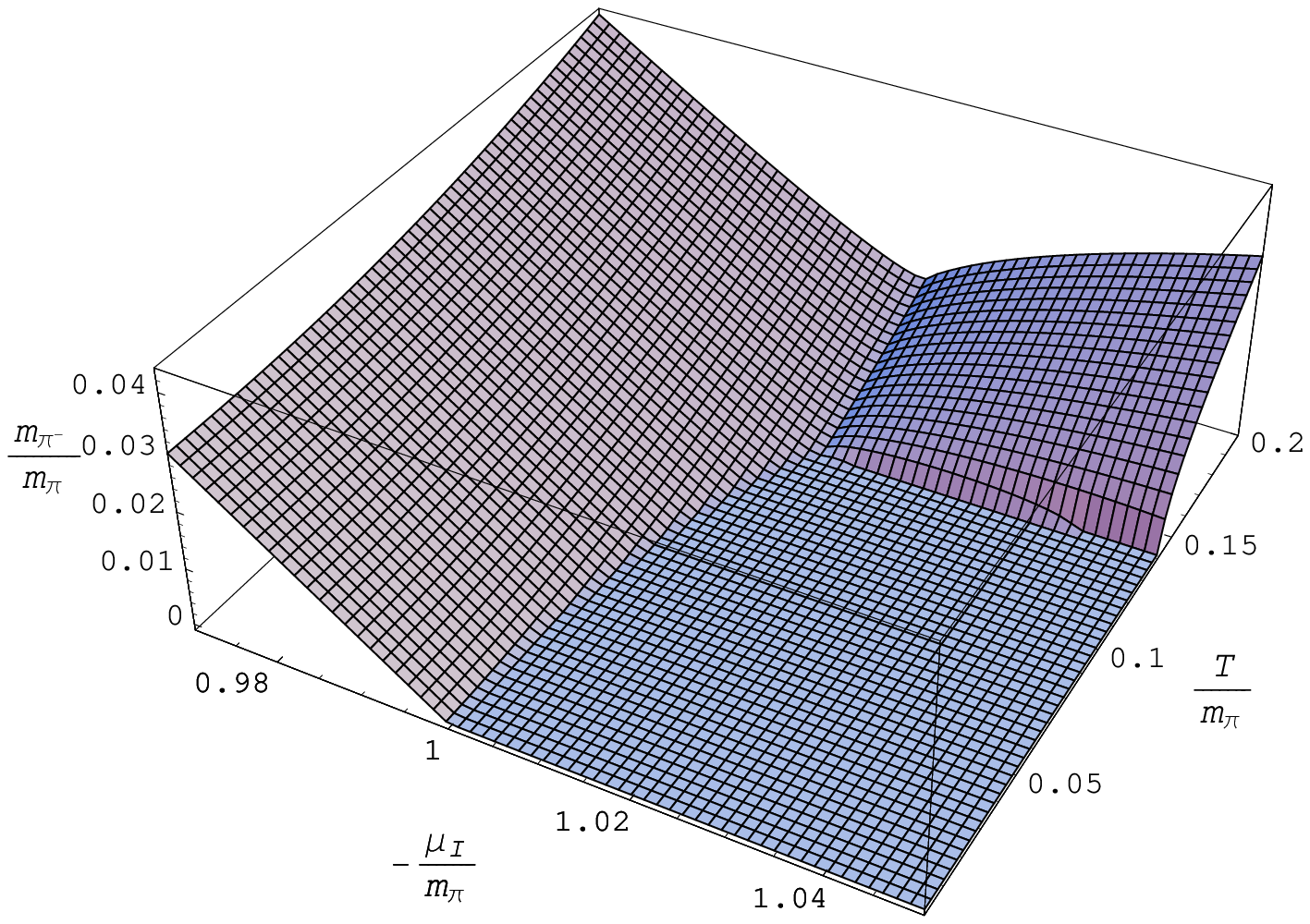}}
\caption{$m_{\pi^-}(T,\mui)$. All the parameters are scaled with
$m_\pi$}
 \label{m-s3D}
\end{figure}

\begin{figure}
 \fbox{
\includegraphics[scale=.62]{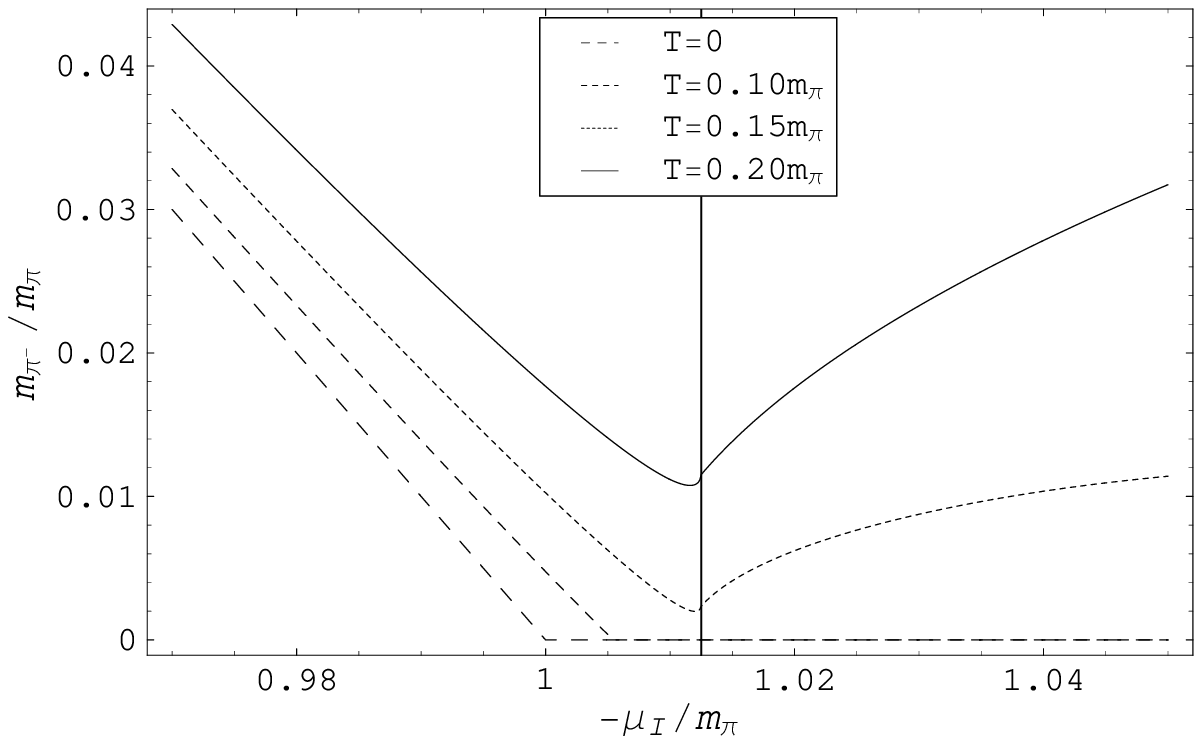}}
\caption{$m_{\pi^-}$ plot versus isospin chemical potential for
different temperatures. All values are scaled with $m_\pi$. The
vertical line corresponds to $\mui = m$.}
 \label{m-s}
\end{figure}

\begin{figure}
\vspace{0cm}
 \fbox{
\includegraphics[scale=.55]{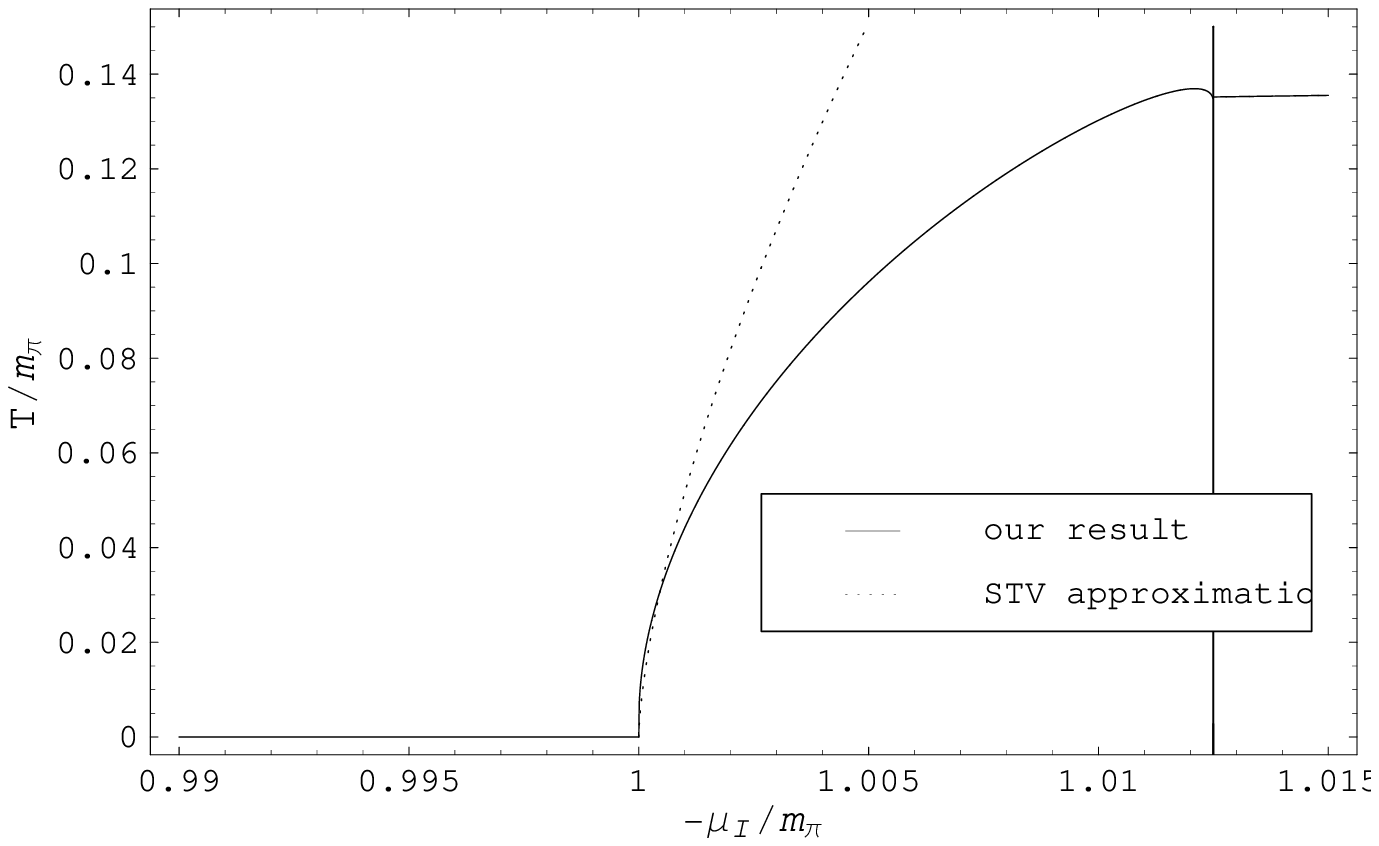}}
\caption{Phase diagram of the condensation point of temperature
versus isospin chemical potential. The dashed line corresponds to
the approximation made by Splittorff, Toublan and Veerbarschot
\cite{Splittorff:2002xn}.  }
 \label{phase}
\end{figure}

\begin{figure}
\vspace{0cm}
 \fbox{
\includegraphics[scale=.7]{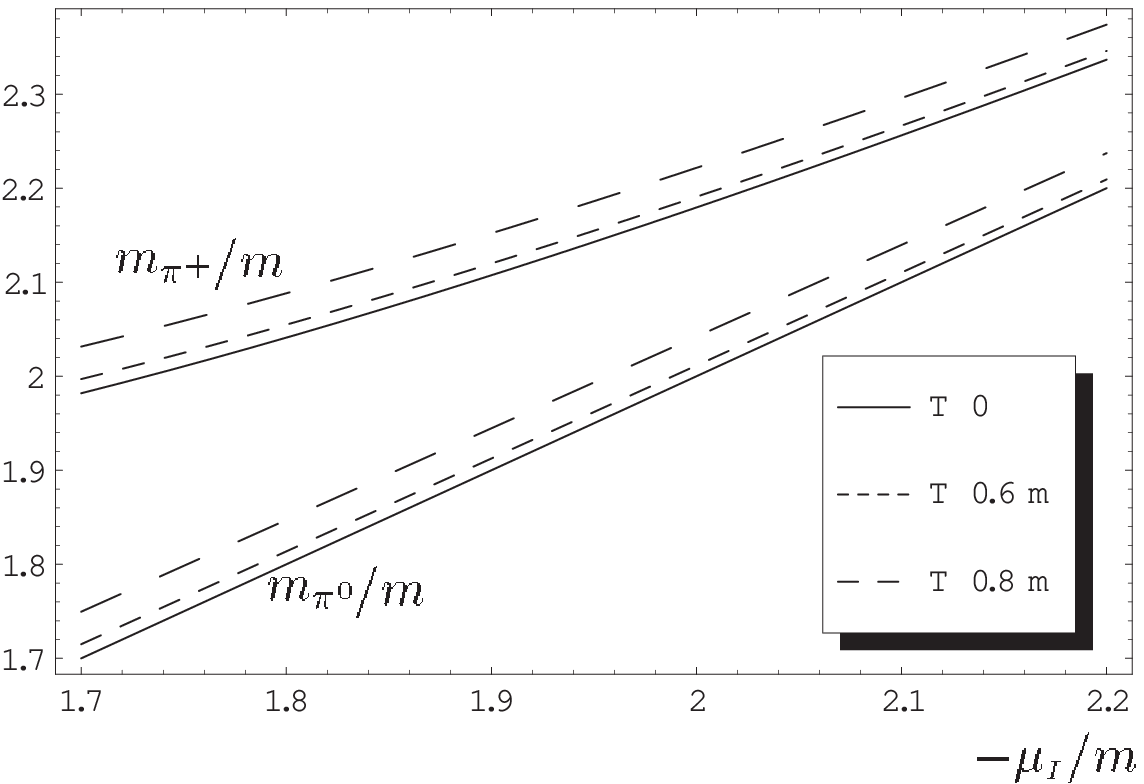}}
\caption{$m_{\pi^0}$ and $m_{\pi^+}$ as a function of high values of the isospin
chemical potential at different values of temperature. All
parameters are scaled with $m$}
 \label{m0mpc}
\end{figure}

To start the discussion, we would like to mention that near the
transition point, $m_{\pi^+}$ decreases, as in the tree-level
approximation, and this behavior is enforced with temperature. In
this region $m_{\pi^0}$ grows with both parameters (see Fig.
\ref{m0mps}).

For $m_{\pi^-}$, according to FIG. \ref{m-s3D} and FIG. \ref{m-s},
we see that a certain critical temperature must be reached before
the $\pi^-$ is removed from the condensed state. This is precisely
the condition that determines the phase transition curve (Fig.
\ref{phase}). In \cite{Loewe:2002tw,Loewe:2003dq},
 we obtained the phase transition
diagram in the $\mui , T$ space, starting from the first phase by
extrapolating the mass evolution $m_\pi(T,\mui)$ up to the point
where it vanishes. This curve has the same shape as the one shown
in 
\cite{Barducci:2004tt} (but for different values, confirming the
discrepancy of about 25\% the authors mentioned, obtained within a
Nambu-Jona-Lasinio model analysis) and indicated previously in 
\cite{Son:2000xc,Son:2000by}.

Starting from the first phase, according to our previous paper
\cite{Loewe:2002tw,Loewe:2003dq}, for $T<m_\pi$, it is possible to find at first order in
$T$ an expression for the transition line $\mui (T)$ that
coincides with the results given in EQ. (8.91) from 
\cite{Splittorff:2002xn}. In
FIG. \ref{phase}, the dashed line corresponds to this
approximation, valid at low temperature. However, as we said, this
result gives us the transition line from the viewpoint of the
first phase. If we start from the condensed phase, considering the
same expansion, except that now $T<|\mui|$ we find that
\begin{equation}
T_{cr}(\mui)\approx|\mui|\left(\frac{\bar
l_3^2}{32\pi\zeta(3/2)^2}\right)^{1/3}
\end{equation}

 The reader could think from FIG.\ref{phase} that the
 critical temperature remains constant in the second phase.
 However, the temperature actually rises as function of
 the chemical potential, as we can see from the previous
 equation, but very slowly,  and therefore we cannot appreciate this
 behavior from the values shown in the figure.
 This growing behavior of the critical temperature,
 is of course consistent with general statements about phase
 transitions in the Ginzburg-Landau theory and it has been
 actually measured in the lattice for two or three color
 QCD.
Nevertheless, if we consider the thermal corrections in the second
phase, it happens that the condensation phenomenon starts to
disappear for a certain value of $T>20MeV$ near the transition
point, remaining a kind of superfluid phase which includes the
condensed as well as the normal phase (massive pion modes).

  In the high chemical potential region, as we said before, the
$\pi^-$ pion condenses again.  For $m_{\pi^0}$ in this region, it
increases monotonically both with temperature and chemical
potential. $m_{\pi^+}$, as the temperature and the chemical
potential rise, becomes asimptotically  close to $m_{\pi^0}$ as
was expected (see FIG. \ref{m0mpc}). In contrast with the first
region, the $\pi^+$ mass grows with temperature and a crossover
occurs somewhere in the intermediate region of the chemical
potential for the temperature dependence, since near the phase
transition point we have $m_{\pi^+}(T_1,\mui)>m_{\pi^+}(T_2,\mui)$
and this behavior changes in the high chemical potential region in
such a way that $m_{\pi^+}(T_1,\mui)<m_{\pi^+}(T_2,\mui)$.

\bigskip
\noindent
 {\bf Acknowledgements:}   The work of  M.L. has been
supported
 by Fondecyt (Chile)
under grant No.1010976. C.V. acknowledges support from  Conicyt

\bibliography{bib}

\end{document}